\documentclass[reprint,pra,twocolumn,twoside,aps]{revtex4-2}

\usepackage{amsmath,amssymb,amsfonts}
\usepackage{graphicx}
\usepackage{siunitx}
\usepackage{natbib}
\usepackage{braket}
\usepackage{multirow}
\usepackage[normalem]{ulem}

\usepackage{url}
\usepackage[breaklinks]{hyperref}
\hypersetup{
    unicode=true,
    colorlinks=true,
    linkcolor=blue,
    citecolor=blue,
    urlcolor=blue
  }
	\usepackage{breakurl}
	\usepackage{xcolor}

\usepackage{xr}

\begin{document}

\title{Quantum Degenerate Mixtures of Cs and Yb}

\author{Kali E. Wilson$^1$}
\thanks{These two authors contributed equally.}
\author{Alexander Guttridge$^1$}
\thanks{These two authors contributed equally.}
\author{Jack Segal$^1$}
\author{Simon L. Cornish$^1$}
\email{s.l.cornish@durham.ac.uk}
\affiliation{$^1$Joint Quantum Centre Durham-Newcastle, Department of Physics, Durham University, South Road, Durham, DH1 3LE, United Kingdom.}

\begin{abstract}
We report the production of quantum degenerate Bose-Bose mixtures of Cs and Yb with both attractive (Cs + $^{174}$Yb) and repulsive (Cs + $^{170}$Yb) interspecies interactions. Dual-species evaporation is performed in a bichromatic optical dipole trap that combines light at 1070\,nm and 532\,nm to enable control of the relative trap depths for Cs and Yb. Maintaining a trap which is shallower for Yb throughout the evaporation leads to highly efficient sympathetic cooling of Cs for both isotopic combinations at magnetic fields close to the Efimov minimum in the Cs three-body recombination rate at around 22\,G. For Cs + $^{174}$Yb, we produce quantum mixtures with typical atom numbers of $N_\mathrm{Yb} \sim 5 \times 10^4$ and $N_\mathrm{Cs} \sim 5 \times 10^3$. We find that the attractive interspecies interaction (characterised by the scattering length $a_\mathrm{CsYb} = -75\,a_0$) is stabilised by the repulsive intraspecies interactions.  For Cs + $^{170}$Yb, we produce quantum mixtures with typical atom numbers of $N_\mathrm{Yb} \sim 4 \times 10^4$, and $N_\mathrm{Cs} \sim 1 \times 10^4$. Here, the repulsive interspecies interaction ($a_\mathrm{CsYb} = 96\,a_0$) can overwhelm the intraspecies interactions, such that the mixture sits in a region of partial miscibility.
\end{abstract}

\maketitle
\section{Introduction}
Mixtures of ultracold atomic gases provide an appealing platform for numerous
avenues of research, including the realisation of novel quantum phases
\cite{Molmer1998,Lewenstein2004a,Zaccanti2006a,Ospelkaus2006a,Guenter2006,Sengupta2007,Marchetti2008}, the formation of quantum droplets \cite{Petrov2015, Cabrera2018, DErrico2019} and solitons \cite{Hamner2011, DeSalvo2019}, the study of collective dynamics \cite{Modugno2002a, Ferrier2014, Roy2017}, binary fluid dynamics and quantum turbulence \cite{Hamner2011, Takeuchi2010}, the investigation of Efimov physics \cite{Tung2014,Pires2014,Maier2015,Ulmanis2016a},
and the creation of ultracold polar molecules
\cite{Ni2008,Molony2014,Takekoshi2014,Park2015,Guo2016, Rvachov2017, Yang2019a,  Voges2020}. 
Early mixture experiments focused on bi-alkali-metal gases
\cite{Modugno2001,Mudrich2002,Hadzibabic2002,Taglieber2008,Spiegelhalder2009,Cho2011,McCarron2011,Wacker2015,Grobner2016}, but there
is currently a growing interest in mixtures composed of alkali-metal and
closed-shell atoms
\cite{Tassy2010,Hara2014,Pasquiou2013,Khramov2014,Vaidya2015,Guttridge2017,Flores2017,Witkowski2017, Ye2020}, as well as mixtures involving the highly magnetic lanthanide atoms Er and Dy \cite{Ravensbergen2018, Trautmann2018, Frye2020}. 
Alkali-alkaline-earth mixtures open up the possibility of creating paramagnetic ground-state
polar molecules, with applications in quantum simulation and quantum
information \cite{Micheli2006,Perez-Rios2010,Herrera2014}, precision
measurement \cite{Alyabyshev2012}, tests of fundamental physics
\cite{Isaev2010,Flambaum2007,Hudson2011}, and tuning of collisions and chemical reactions \cite{Abrahamsson2007,Quemener2016}.  
The high phase space densities (PSD) afforded by dual-degenerate mixtures will improve prospects for magnetoassociation via  the relatively inaccessible interspecies Feshbach resonances  between alkali and alkaline-earth-like atoms \cite{Zuchowski2010, Brue2012, Brue2013, Barbe2018, Yang2019, Green2020}. 

In general, the mean-field properties of a quantum mixture are dictated by the relative magnitude and signs of the intraspecies and interspecies interactions, which are characterised by the two intraspecies scattering lengths and the single interspecies scattering length. Varying the relative strengths of these interactions allows one to tune the collective dynamics of the system, opening up exploration of beyond-mean field effects such as quantum fluctuations \cite{Lee1957, Petrov2015}. Cs--Yb mixtures offer several advantages in this context.  The rich Feshbach structure of Cs gives precise control of the Cs intraspecies scattering length $a_\mathrm{Cs}$ at low magnetic field \cite{Chin2000}; $a_\mathrm{Cs}$ can be tuned without affecting Yb, which has no electronic magnetic moment in its $^1 S_0$ ground state.  Yb has seven stable isotopes (five bosonic and two fermionic),  enabling the study of both Bose--Bose and Bose--Fermi systems.  To date all but the bosonic isotope $^{172}$Yb have been cooled to degeneracy \cite{Takasu2003,  Fukuhara2007a, Fukuhara2007b, Fukuhara2009, Taie2010, Sugawa2011}. Switching between Yb isotopes allows for a degree of tunability of the Yb intraspecies scattering length \cite{Kitagawa2008}.  Moreover, the large mass of Cs combined with the number of Yb isotopes available, results in a large range of tunability for the atom-pair reduced mass, and hence the Cs--Yb interspecies scattering length.  Recent photoassociation experiments \cite{Guttridge2018a} were used to constrain the theoretical model of the CsYb molecular ground state interaction potential, leading to accurate calculations of both the interspecies scattering lengths \cite{Guttridge2018a} and the locations of interspecies Feshbach resonances \cite{Yang2019}.  Notably, Cs and $^{174}$Yb have an attractive interspecies scattering length $a_\mathrm{CsYb} = -75(3)\, a_0$ on the same order of magnitude as the repulsive Yb intraspecies scattering length $a_\mathrm{Yb} = 105(2) \, a_0$ \cite{Kitagawa2008}, ideal for studies of quantum droplets and beyond mean-field physics \cite{Petrov2015}. In contrast, Cs and $^{170}$Yb have a repulsive interspecies scattering length $a_\mathrm{CsYb} = 96.2(2)\, a_0$ and a repulsive Yb intraspecies scattering length $a_\mathrm{Yb} = 64(2) \, a_0$ \cite{Kitagawa2008}, suitable for miscibility studies and binary fluid dynamics.  Varying the Cs scattering length thus enables the balance of interspecies and intraspecies interactions to be tuned across a miscible/immiscible or Bose-Einstein condensate (BEC)/droplet phase transition. We note that the mass scaling of the Cs--Yb interspecies scattering length is similar to that of Rb--Yb (including both $^{85}$Rb and $^{87}$Rb isotopes) \cite{Borkowski2013}. The primary advantages to using Cs as the alkali atom lie in the fine control of the Cs intraspecies scattering length afforded by the shallow slope of the  scattering length versus magnetic field $da/dB \sim 65\,a_0/\mathrm{G}$ in the region spanning the zero crossing at 17\,G \cite{Berninger2013}, and the experimental accessibility of magnetic Feshbach resonances at lower magnetic fields \cite{Chin2000, Marte2002, Blackley2013}. This is particularly relevant for studies of droplet physics, which require both an attractive interspecies interaction and fine control of the balance of mean-field interactions.

The different electronic structures inherent  in  quantum  mixtures  formed  of  alkali  and  alkaline-earth-like  atoms, such as Cs and Yb, present additional advantages. Their distinct structure facilitates trapping in species-specific optical potentials with photon scattering rates much lower than mixtures of alkali atoms \cite{Tassy2010, Herold2012, Vaidya2015}. The confinement of atomic mixtures in these species-specific potentials enables studies of impurity physics and topological superfluids in mixed dimensions \cite{Wu2016, Loft2017, Schafer2018}.
Further species-specific manipulation of the mixture with magnetic forces is possible, as Yb's filled outer shell means that there is no magnetic moment arising from electronic spin for Yb in its $^1 S_0$ ground state.  Additionally, bosonic Yb isotopes have no nuclear spin.
The fermionic isotopes $^{171}$Yb and $^{173}$Yb have nuclear spins of I = 1/2 and I = 5/2, respectively; this additional nuclear degree of freedom has  previously been used to access SU(N) physics \cite{Taie2010}.  The ability to independently perturb each atomic species will enhance studies of collective dynamics \cite{Modugno2002a, Ferrier2014, Roy2017, DeSalvo2019}, and vortex interactions in quantum mixtures \cite{Yao2016, Kuopanportti2018}. {It is important to note that t}he differences in atomic structure also lead to some experimental challenges.  In particular, the different atomic polarizabilties make it challenging to create a suitable optical trap for both species using a single wavelength, thus motivating our use of a bichromatic optical dipole trap (BODT). 

In this paper we report quantum degenerate mixtures of Cs and Yb.  We have produced dual-degenerate Bose-Bose mixtures of Cs--$^{174}$Yb and Cs--$^{170}$Yb in a BODT. The BODT combines light at 1070\,nm and 532\,nm to allow us to dynamically control the relative trap depths for Cs and Yb. For both mixtures, Cs is sympathetically cooled by the Yb atoms, providing an efficient route to dual-degeneracy. The paper is organized as follows: Section~\ref{sec:exp} gives an overview of our experiment and previously developed pathways to single-species Cs and Yb BECs. Section~\ref{sec:BODT} discusses the challenges to dual-species evaporation of Cs and Yb, and motivates the need for a BODT. The experimental implementation of the BODT is discussed in Sec.~\ref{sec:BODT1}. Section~\ref{sec:BODT2} describes the sequential preparation of Cs and Yb samples in the BODT. Section~\ref{sec:evap} covers the dual-species evaporation sequence.  Section~\ref{sec:BODT2} and Sec.~\ref{sec:evap} focus on the attractive Cs--$^{174}$Yb mixture.  Section~\ref{sec:170Yb} extends dual-species evaporation to a Cs--$^{170}$Yb mixture, and explores the role of the repulsive interspecies interaction.  Section~\ref{sec:conclusion} concludes the article.

\section{Experimental Overview: production of single species BECs}\label{sec:exp}
 Our dual-evaporation method builds upon our previous experience with creating single-species BECs of both Cs and $^{174}$Yb  \cite{Guttridge2017}. The different physical properties of the two species leads to very different pathways to degeneracy with conflicting optical trap requirements.  The experimental details have been previously reported elsewhere \cite{Hopkins2016, Kemp2016, Guttridge2016,  Guttridge2017}; here we briefly summarise the relevant details. 

We employ a dual-species oven to generate atomic beams of Cs and Yb which are slowed to their respective magneto-optical-trap (MOT) capture velocities using a dual-species Zeeman slower \cite{Hopkins2016, Kemp2016}. For Yb, the Zeeman slower operates on the broad $^1 S_0$ to $^1 P_1$ transition at 399\,nm (linewidth $\Gamma/2\pi = 29 \, \mathrm{MHz}$).  The atoms are loaded directly into a MOT operating on the $^1 S_0$ to $^3 P_1$ transition at 556\,nm $\Gamma/2\pi = 182 \, \mathrm{kHz}$) \cite{Guttridge2016}.  The narrow linewidth of the intercombination line results in a low Doppler cooling limit of 4.4\,$\mu$K. We load $\sim 2 \times 10^7$ $^{174}$Yb atoms directly from the MOT into a high-power  $1070\,\mathrm{nm}$ crossed optical dipole trap (ODT) with waists of $33(3)\, \mu \mathrm{m}$ and $72(4)\, \mu \mathrm{m}$.  We use a total power of 55 W, resulting in an initial trap depth of $U_\mathrm{Yb} = 950 \, \mu \mathrm{K}$. We perform forced evaporative cooling to degeneracy (critical temperature $T_\mathrm{C,Yb} \approx 350 \, \mathrm{nK}$) by reducing the total power in the ODT to $\sim 1\,\mathrm{W}$ in an approximately exponential ramp.  The low three-body loss rate for $^{174}$Yb combined with the favorable scattering length $a_\mathrm{Yb} = 105 \, a_0$ means high-density and  high collision rates can be used to make evaporation fast and efficient.  We typically create pure $^{174}$Yb BECs containing up to $4 \times 10^5$ atoms.

In contrast, when cooling Cs to degeneracy, the less favorable Cs scattering properties mean that much greater care is required when managing the trap parameters.  In particular, it is critical to manage the density of the Cs sample to mitigate the strong inelastic losses due to three-body recombination \cite{Weber2003}; the Cs three-body loss rate is $K_3 \sim 1 \times 10^{-27}\,\mathrm{cm}^{6}/\mathrm{s}$ \cite{Kraemer2006,Guttridge2017}. For Cs, both the Zeeman slower and the MOT operate on the $^2 S_{1/2}$ to $^2 P_{3/2}$ transition at 852\,nm (${\Gamma}/2\pi = 5.23\, \mathrm{MHz}$).  In contrast to the narrow-line Yb MOT, the Doppler limit for the Cs MOT is $125 \, \mu \mathrm{K}$, requiring an additional cooling stage prior to loading into the ODT. Our approach follows the method developed in Innsbruck \cite{Weber2003a, Kraemer2004}.
We use optical molasses followed by degenerate Raman sideband cooling (DRSC) to precool the Cs sample to $T_\mathrm{Cs} \sim 2 \, \mu\mathrm{K}$.  The DRSC simultaneously optically pumps the atoms to the $\ket{F = 3, m_\mathrm{F} = + 3}$ hyperfine state. We then load the atoms into a large-volume levitated 1070\,nm ODT referred to as the reservoir. Approximately $10 \%$ of the atoms confined in the reservoir are then loaded into a tightly focused ODT, referred to as the dimple. The atoms confined in the dimple trap experience negligible heating, despite the increased density, as they are cooled through collisions with the bath of reservoir atoms. The dimple trap uses the same 1070\,nm beams as the ODT used for Yb evaporation.  The total power in the dimple trap is 280\,mW, resulting in an initial trap depth $U_\mathrm{Cs} =  20\, \mu \mathrm{K}$, much weaker than the initial Yb trap.  Forced evaporation is performed by reducing the power in the dimple beams to approximately $20\, \mathrm{mW}$ allowing Cs to cross the BEC transition at $T_\mathrm{C,Cs} \approx 60 \, \mathrm{nK}$.  During the evaporation process we use a bias field of 22.3 G to place the Cs intraspecies scattering length in the vicinity of the Efimov minimum in the three-body recombination rate \cite{Kraemer2006}, and thus optimize the ratio of elastic to inelastic collisions \cite{Weber2003}.  We typically create pure Cs BECs containing up to $5 \times 10^4$ atoms in the $\ket{F = 3, m_\mathrm{F} = +3}$ hyperfine state.

\section{Motivation for a bichromatic optical dipole trap (BODT)}\label{sec:BODT}

The primary challenge which must be overcome to produce a degenerate Cs--Yb mixture is the very different traps required for efficient evaporation. The optimal traps for single-species evaporation described in Sec.~\ref{sec:exp} are incompatible.  The powers required for the final Yb trap are much greater than the powers required for loading Cs ($P_\mathrm{f, Yb} \sim 4 P_\mathrm{i, Cs}$), and at the final power needed for Cs the trap is non-existent for Yb ($P_\mathrm{f, Cs} \sim 0.02 P_\mathrm{f, Yb}$). Here $P_{i,\alpha}$ denotes the power required to load a given species $\alpha = \{\mathrm{Cs}, \mathrm{Yb}\}$, while $P_{f,\alpha}$ denotes the power at the end of the given single-species evaporation sequence. This incompatibility is largely due to the different atomic polarizabilites of the two species.  At the ODT wavelength $\lambda = 1070 \, \mathrm{nm}$, the magnitude of the polarizability for Cs is over seven times that for Yb resulting in a trap which is always deeper and tighter for Cs than for Yb. 

We therefore need a way of tuning both absolute and relative trap depths over a large dynamic range.  Ideally, we want a trap which is marginally deeper for Cs so that the Cs atoms can be sympathetically cooled by Yb.  However, we also need the Cs trap to be sufficiently shallow to avoid atom loss arising from  three-body recombination. Lastly, we want to be able to dynamically tune the relative trap depths throughout the evaporation process. This need to control the relative Cs and Yb trap depths highlights the need for an alternative to the single wavelength traps at 1064\,nm or 1070\,nm typically employed in ultracold atom experiments.  

\begin{figure}[t]
		\includegraphics[width=0.95\linewidth]{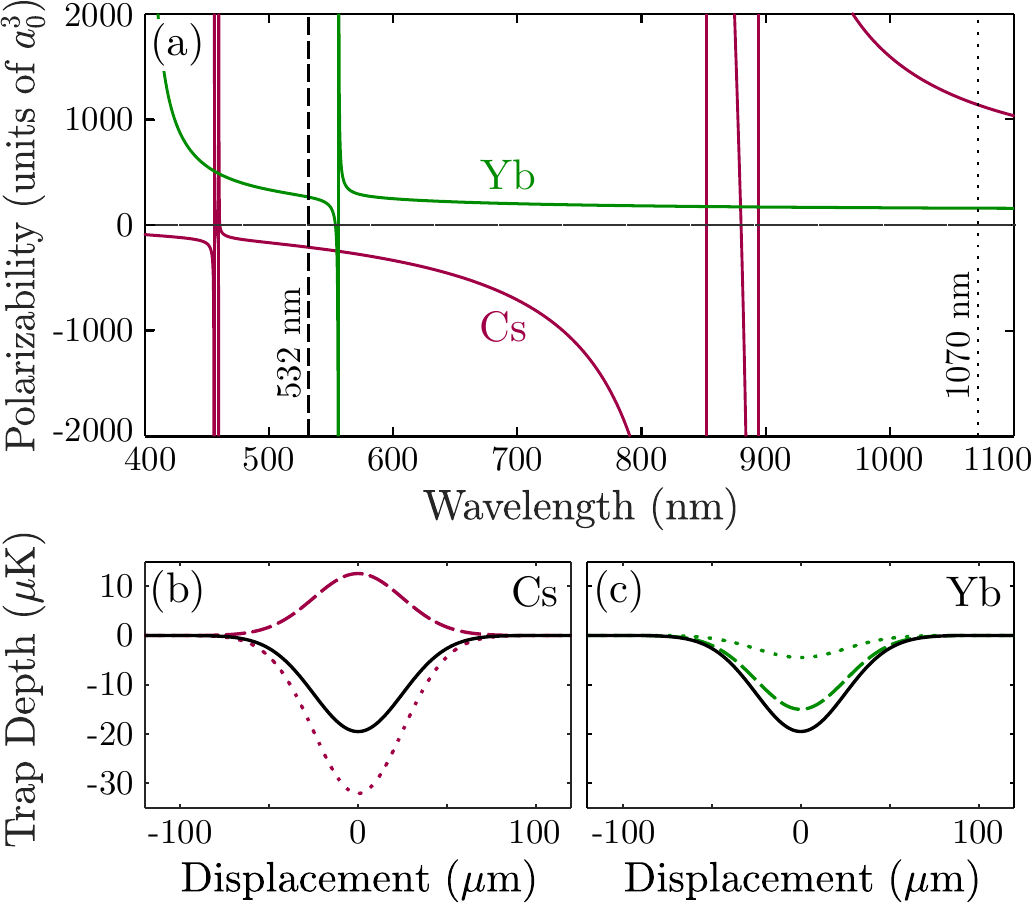}
\caption{Balanced trapping potentials for Cs and Yb. (a) Atomic polarizability versus wavelength for Cs (red) and Yb (green).  For reference, vertical lines are placed at 532\,nm (dashed) and 1070\,nm (dotted), the wavelengths of the bichromatic optical dipole trap (BODT) beams. The polarizability of Cs (Yb) is $-211\,a_0^3$ ($264\,a_0^3$) at 532\,nm and $1142\,a_0^3$ ($160\,a_0^3$) at 1070\,nm.  (b) and (c) Cross sections of a representative balanced trapping potential formed from the sum of the 532\,nm (dashed) and 1070\,nm (dotted) potentials for Cs and Yb, respectively. Both beams have $1/e^2$ waists of $w_0 = 50\,\mu\mathrm{m}$ and the powers are $P_{532} = 1\,\mathrm{W}$ and $P_{1070} = 0.5\,\mathrm{W}$.
	\label{fig:pol}}
\end{figure}

We consider two strategies to producing a trap suitable for simultaneous evaporation of Cs and Yb. The first involves tuning the wavelength of the ODT to a value where the relative polarizabilities are highly tunable, i.e., a `special-wavelength' ODT. For Cs--Yb the most promising region is near the Cs $6^2S_{1/2} - 7^2P_{1/2}$ and $6^2S_{1/2} - 7^2P_{3/2}$ doublet transitions where the Cs polarizability changes rapidly with wavelength, while the Yb polarizability varies slowly, as shown in Fig.~\ref{fig:pol}(a). We note that the region between the more familiar Cs D1 and D2 lines also fulfills this criteria. However, the region near 460\,nm is preferable as the narrow linewidths of the Cs doublet transitions (more than a factor of 10 narrower than the D1 and D2 lines), and the larger Yb polarizability ($\sim 500\,a_0^3$ near 460\,nm compared to $\sim 170\,a_0^3$ near 880\,nm) mean that Cs experiences less heating for a given Yb trapping potential. For example, the two species experience a trap with equal depth at $\lambda = 456.05 \, \mathrm{nm}$, and a trap with equal frequencies $\omega \propto \sqrt{\alpha / m}$ at $\lambda = 456.14 \, \mathrm{nm}$, where $m$ is the atomic mass, and $\alpha$ is the polarizability. Here we have used the atomic mass for $^{174}$Yb to calculate the wavelength that gives equal trapping frequencies.   

However, it is both challenging and expensive to generate sufficient optical power at these wavelengths.   Alternatively, we can employ a BODT \cite{Tassy2010,Vaidya2015, Onofrio2002} using laser beams of two different wavelengths, where one of the wavelengths has opposite sign polarizabilities for the two species. This allows standard wavelengths to be used where there is a lot of available power, e.g., $532 \,$nm and 1070\,nm.  As shown in Fig.~\ref{fig:pol}(a), Cs has a negative polarizability at $532 \, \mathrm{nm}$  [$\alpha_\mathrm{Cs} = -211\,a_0^3$], which can be used to reduce the large trapping potential experienced by the Cs atoms due to the large positive polarizability at $1070 \, \mathrm{nm}$  [$\alpha_\mathrm{Cs} = 1142\,a_0^3$]. In contrast, Yb is trapped at both wavelengths, with $\alpha_\mathrm{Yb} = 264\,a_0^3$ at 532\,nm and $\alpha_\mathrm{Yb} = 160\,a_0^3$ at 1070\,nm. Varying the power in the $532 \,$nm trapping beam relative to the $1070 \, $nm beam thus allows the ratio of trap depths for the two species to be tuned to a suitable value throughout the dual-species evaporation.  A representative example of a balanced trapping potential is shown in Fig.~\ref{fig:pol}(b) and (c) for Cs and Yb, respectively. 

\section{Implementation of the BODT}\label{sec:BODT1}
\begin{figure}[t]
		\includegraphics[width=0.95\linewidth]{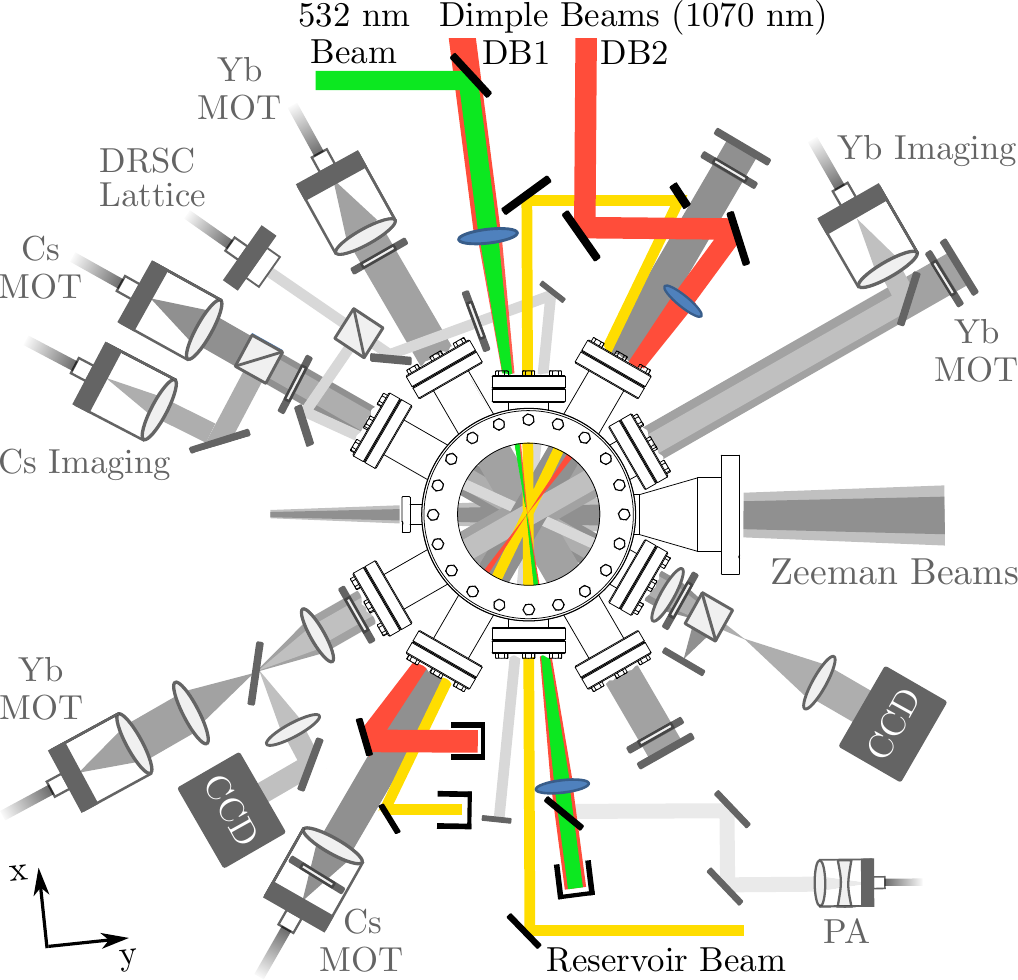}
\caption{Optical beam layout for the bichromatic optical dipole trap (BODT) overlaid on top of a vertical view of all beam paths traveling through the science chamber.   The BODT consists of a $532 \, \mathrm{nm}$ beam copropagating with a $1070 \, \mathrm{nm}$ beam (DB1), which are arranged at a $40^\circ$ angle to the second $1070 \, \mathrm{nm}$ beam (DB2). We define the coordinate system ($x,y,z$) with ${x}$ along the copropagating beams, ${y}$ perpendicular to ${x}$ in the horizontal plane, and ${z}$ the vertical direction. The Cs (Yb) imaging axes are rotated $35^{\circ}$ ($25^\circ$) from the ${y}$ axis.
	\label{fig:beams}}
\end{figure}
\begin{figure*}[ht]
		\includegraphics[width=0.95\linewidth]{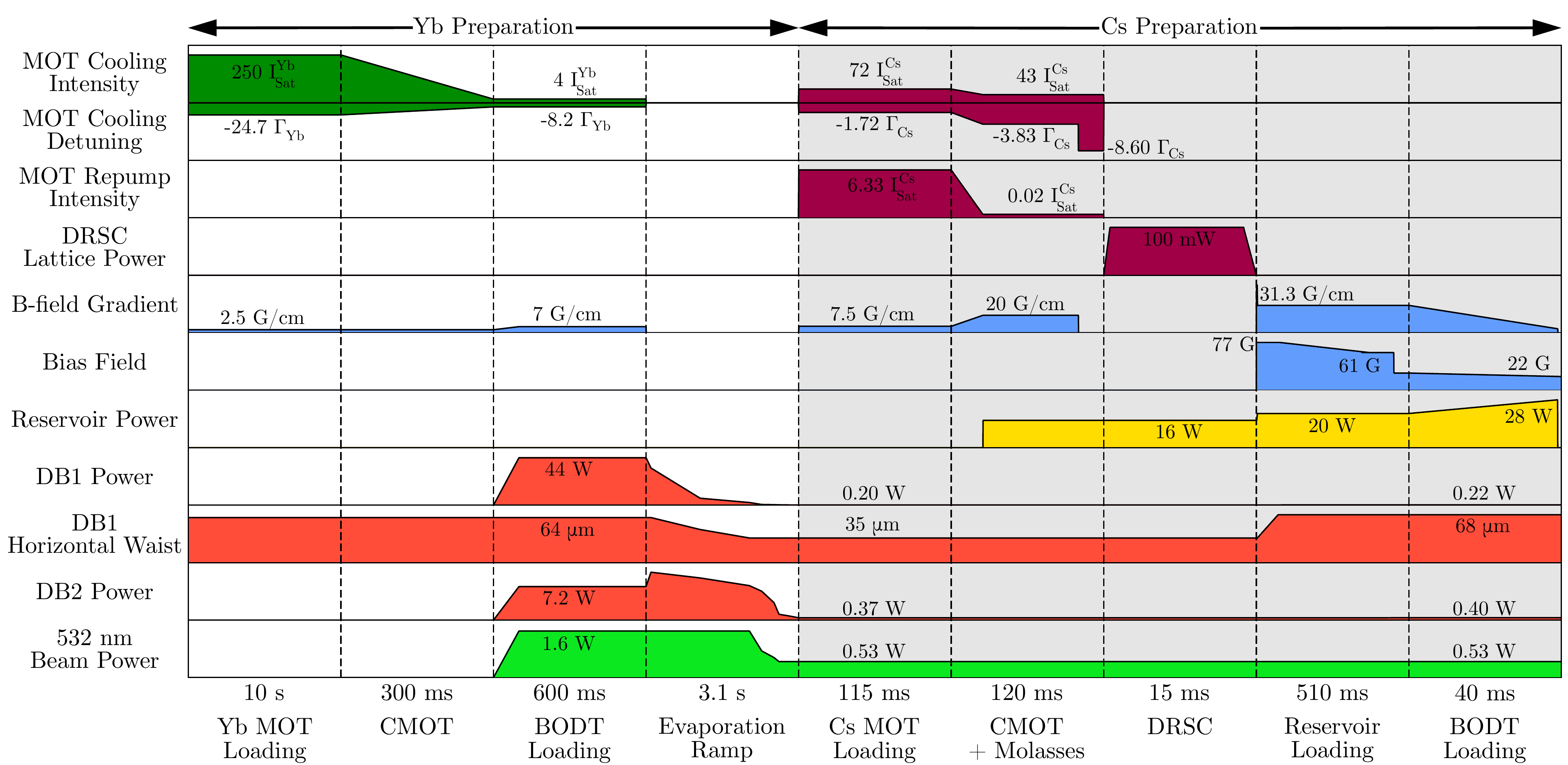}
\caption{Timing sequence for Cs and Yb preparation in the bichromatic optical dipole trap (BODT). Yb is loaded in a magneto-optical trap (MOT) which is then compressed (CMOT) and loaded into the BODT. The Yb MOT operates on the $^1 S_0$ to $^3 P_1$ transition at 556 $\,$nm; linewidth ${\Gamma}_{\mathrm{Yb}}/2\pi=182.2\,$kHz and saturation intensity ${\mathrm{I}}_{\mathrm{Sat}}^{\mathrm{Yb}}=0.139\,\mathrm{mW}\,{\mathrm{cm}}^{-2}$. Yb undergoes an initial 3.1 s of forced evaporation. Cs is then loaded in a MOT and compressed (CMOT), followed by optical molasses, and degenerate Raman sideband cooling (DRSC). The Cs MOT operates on the $^2 S_{1/2}$ to $^2 P_{3/2}$ transition at 852\,nm; ${\Gamma}_{\mathrm{Cs}}/2\pi=5.23\,$MHz and ${\mathrm{I}}_{\mathrm{Sat}}^{\mathrm{Cs}}=1.105\,\mathrm{mW}\,{\mathrm{cm}}^{-2}$. The MOT cooling beams are detuned from the $F=4$ to $F'=5$ transition as shown, and the MOT repump beams are detuned from the $F=3$ to $F'=4$ transition by a constant -5.4\,MHz. The DRSC lattice beams are resonant with the $F=4$ to $F'=4$ transition. After DRSC, Cs is loaded into a hybrid optical and magnetic reservoir trap, and from there atoms are loaded into the BODT. The BODT consists of two 1070$\,$nm dimple beams (DB1 and DB2), and one 532$\,$nm beam. DB1's horizontal waist is controlled by dithering the frequency of the AOM used to servo DB1's power, with a range of 35 to 68$\,\mu$m. DB1's vertical waist is 29(2)$\,\mu$m. The beam waists of DB2 and the 532$\,$nm beam are 70(4)$\,\mu$m and 50(3)$\,\mu$m respectively. To reduce heating of Cs during the reservoir load, we use slightly lower powers of DB1 (0.2 W) and DB2 (0.37 W) during the Cs MOT and reservoir load compared to the final powers for the Cs BODT load.  
	\label{fig:timing}}
\end{figure*}

In this article, we pursue the second strategy discussed in Sec.~\ref{sec:BODT}, employing a BODT to create a trapping potential suitable for cooling Cs--Yb mixtures to quantum degeneracy. The geometry of the BODT is illustrated in Fig.~\ref{fig:beams}, which gives a top-down view of the science chamber showing the alignment of the BODT beams with respect to the other laser beams traveling through the chamber. Here we define the coordinate system $(x,y,z)$ with $x$ along the copropagating beams, $y$ perpendicular to $x$ in the horizontal plane, and $z$ the vertical direction. We note that the Cs (Yb) imaging axes are rotated $35^\circ$ ($25^\circ$) from the $y$ direction.

The BODT trapping configuration consists of two $1070\,$nm beams, referred to as dimple beam 1 (DB1) and dimple beam 2 (DB2), crossed at an angle of $40 ^{\circ}$. This dimple beam geometry is the same as that used for our single species evaporation schemes. The $532 \,$nm light is combined with the 1070\,nm light using a dichroic mirror and copropagates with DB1.  The 532\,nm beam and DB2 have a symmetric beam waists of $50(3) \, \mu $m and $70(4) \, \mu$m, respectively.  DB1 has horizontal and vertical $1/e^2$ beam waists of $w_y = 29(2) \, \mu$m and $w_z = 35(3) \, \mu$m. We vary the horizontal waist of DB1 by dithering the frequency of the AOM used to servo DB1's power; this creates an approximately Gaussian time averaged potential with a variable horizontal waist and gives an additional degree of freedom over the trap frequencies \cite{Ahmadi2005}. This technique has been employed in other similar experiments \cite{Trautmann2018, Baier2012, Tang2015, Roy2016b}.

The relative alignment of the copropagating BODT beams is critical to maintain both optimal trap depths and good overlap between the two atomic species. Due to the different polarizabilities, Cs is more strongly trapped by DB1, and Yb by the 532\,nm beam which can lead to strange trapping potentials, e.g., a double well potential for Cs, if the beams are not well aligned (one of the disadvantages of BODTs). With this in mind, we track the relative alignment of the copropagating beams with a quadrant photodiode (QPD), which gives the relative beam positions at the location of the atoms to within 3\,$\mu$m in both $y$ and $z$ directions. We use Yb atoms confined in a single beam to both calibrate the QPD and verify coarse alignment of the BODT beams.  Further verification of optimal beam alignment is determined by the thermalization measurements discussed in Sec.~\ref{sec:BODT2}. The axial $x$ alignment of the focused beam waists is less critical as the 532\,nm beam has a Rayleigh range of 15\,mm. Moreover, confinement in the axial direction is predominately due to DB2. Further information regarding the QPD setup and calibration is given in the Appendix.

\section{Preparing Cs and Yb in the BODT}\label{sec:BODT2}
We load Yb and Cs sequentially into the BODT using the timing sequence shown in Fig.~\ref{fig:timing}. Following the approaches used in the single-species cooling schemes outlined in Sec.~\ref{sec:exp}, we begin by preparing $^{174}$Yb in the BODT. We load Yb first because of the long MOT load time (typically $10-20\,\mathrm{s}$), and the initial stage of evaporative cooling requires a tight, high power trap. Together these factors would cause intolerable heating and loss of Cs if it were loaded first. As shown in Fig.~\ref{fig:timing}, the BODT potential is initially dominated by the $1070 \,$nm trapping light, with the horizontal beam waist of DB1 set to $64\,\mu\mathrm{m}$ to increase the trap volume for better loading. We perform forced evaporation of Yb by reducing the power of all three BODT beams, resulting in an approximately exponential ramp of the Yb trap depth over 3.1 s. We simultaneously decrease the horizontal waist of DB1, which partially combats the detrimental drop in trap frequency associated with reducing the beam powers.

We then begin the preparation of the Cs sample following the method outlined in Sec.\,\ref{sec:exp}. The Cs MOT is loaded, and then compressed (CMOT) by reducing the intensity and detuning of the MOT beams.  The atoms are further cooled using optical molasses, followed by DRSC, which cools the atoms to  $\approx 1 \, \mu$K and simultaneously optically pumps them into the $\ket{F=3, m_\mathrm{F}=+3}$ hyperfine state. To enhance Cs loading into the BODT we transfer the remaining atoms into a large-volume hybrid optical and magnetic reservoir trap. To efficiently capture the atoms in the reservoir, we modify the elastic collision rate by varying the magnetic field during the loading and apply a brief force vertically using a magnetic field gradient of 55\,G/cm, before reducing the gradient to 31.3\,G/cm to levitate the atoms \cite{Li2015}. While loading into the reservoir we also increase the horizontal beam waist of DB1 to $68 \, \mu\mathrm{m}$. The extra degree of freedom afforded by varying DB1's waist allows us to create a suitable trapping potential for the BODT loading and subsequent sympathetic cooling of Cs atoms by reducing the mean trap frequency of the BODT to $\bar{\omega}_\mathrm{Cs}/2\pi = 123 \, \mathrm{Hz}$, and the Cs trapping potential to $U_\mathrm{Cs} = 11\, \mu \mathrm{K}$. It also slightly reduces the Yb trapping potential to $U_\mathrm{Yb} = 3.1 \, \mu \mathrm{K}$, with a mean trap frequency of $\bar{\omega}_\mathrm{Yb}/2\pi = 93 \, \mathrm{Hz}$.
After 510\,ms of reservoir loading, we ramp down the magnetic field gradient to 0\,G/cm, and then extinguish the reservoir trapping light, leaving a fraction of the reservoir atoms confined in the BODT colocated with the Yb sample. This is the starting point for the dual-evaporation sequence discussed in Sec.~\ref{sec:evap}.  Preparing the Cs sample in the presence of Yb takes $\sim 1\,\mathrm{s}$,  5-10 times faster than the preparation time required for Cs on its own. This is largely because our strategy relies on sympathetic cooling which is incredibly efficient for the species being cooled and, hence, we do not need to load many Cs atoms. In fact, loading too many Cs atoms is detrimental as it places too much of a heat load on the Yb atoms. 
\begin{figure}[t]
		\includegraphics[width=0.95\linewidth]{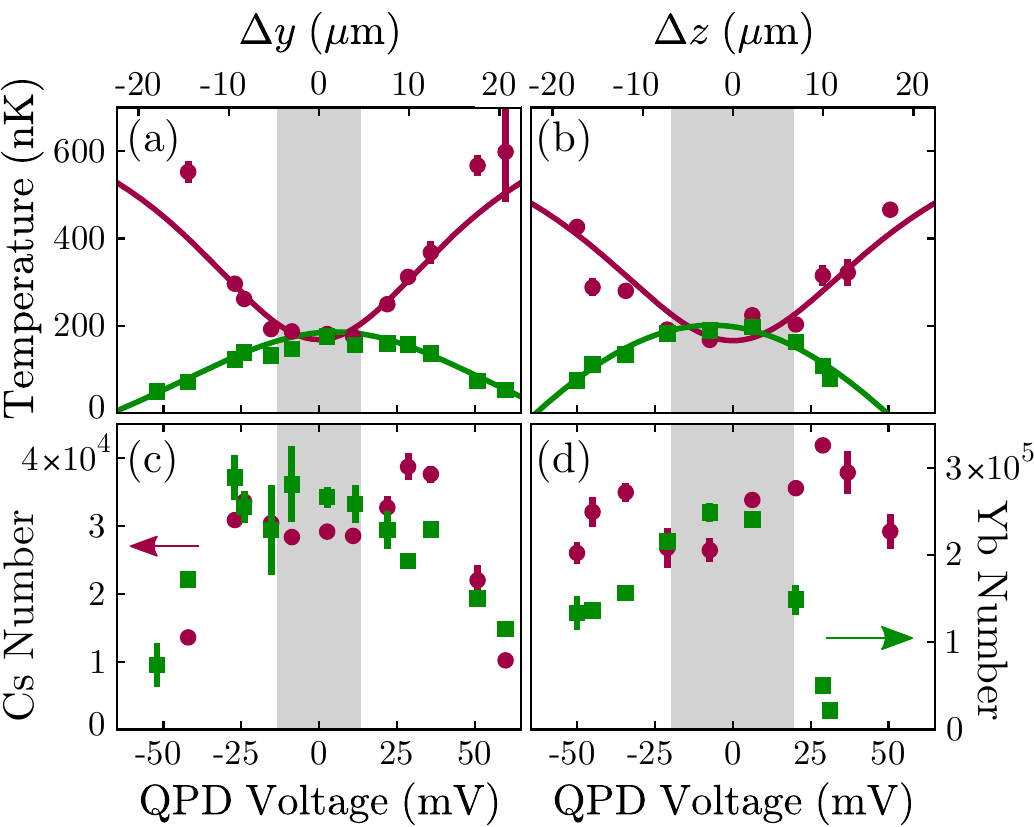}
\caption{Optimal atom cloud overlap indicated by increased thermalization as a function of the 532\,nm beam displacement measured with a quadrant photodiode (QPD). (a) and (b) The Cs (red circles) and Yb (green squares) temperatures versus beam displacement for (a) horizontal displacement $\Delta y$, and (b) vertical displacement $\Delta z$.  We define zero to be the beam position corresponding to the minimum temperature for the Cs atoms.  Solid lines are Lorentzian fits to the temperature data.  (c) and (d) The Cs (red circles) and Yb (green squares) atom numbers versus beam displacement.  Grey shading corresponds to the regions where the atoms are in good thermal contact $T_\mathrm{Cs} \sim T_\mathrm{Yb}$.  
\label{fig:thermal_QPD}}
\end{figure}

Maintaining good overlap of the atomic samples is critical to efficient sympathetic cooling of Cs, and requires careful alignment of the BODT beams.  Figure~\ref{fig:thermal_QPD} shows the temperature and atom numbers for the Cs and Yb samples confined in the BODT versus $532\,\mathrm{nm}$ beam displacement.  The measurements are taken midway through the dual-evaporation sequence. We track the vertical $\Delta z$ and horizontal $\Delta y$ 532-nm beam displacement with the QPD, and define zero to be the position corresponding to the minimum Cs temperature. As expected, the lowest Cs temperature coincides with the highest Yb temperature due to the thermal contact between the cold bath of Yb and the hotter Cs. Therefore, minimising the Cs temperature is a signature of good overlap between the atomic clouds.  As shown in Fig.~\ref{fig:thermal_QPD}(a) and Fig.~\ref{fig:thermal_QPD}(c), the measured temperature and atom numbers are symmetric about $\Delta y = 0 \, \mu \mathrm{m}$. 
In contrast, Fig.~\ref{fig:thermal_QPD}(d) reveals a vertical asymmetry for the Yb atom number as a function of $\Delta z$. In the vertical direction, the alignment for best thermal contact likely does not correspond to the 532\,nm beam and DB1 being colocated vertically as the Yb atoms undergo a significant amount of gravitational sag due to the weaker Yb vertical trap frequency. The lighter Cs atoms experience a tighter trap such that sag is reduced. Therefore, good thermal contact between the Cs and Yb clouds at the end of dual-evaporation requires the 532\,nm beam to be $5-10 \, \mu \mathrm{m}$ above DB1, shifting the minimum of the net Yb trapping potential upwards with respect to Cs.  However, moving the 532\,nm beam too far above DB1 results in a loss of all Yb atoms due to reduction in the trap depth of the weak Yb trap.
We have also confirmed the overlap of the Cs and Yb atoms by using photoassociation loss spectroscopy \cite{Guttridge2018}, measuring the reduction in detected Cs atoms after exciting the atomic mixture to an excited molecular state CsYb*.

\section{Evaporation to dual-degeneracy}\label{sec:evap}
\begin{figure}[t]
		\includegraphics[width=0.95\linewidth]{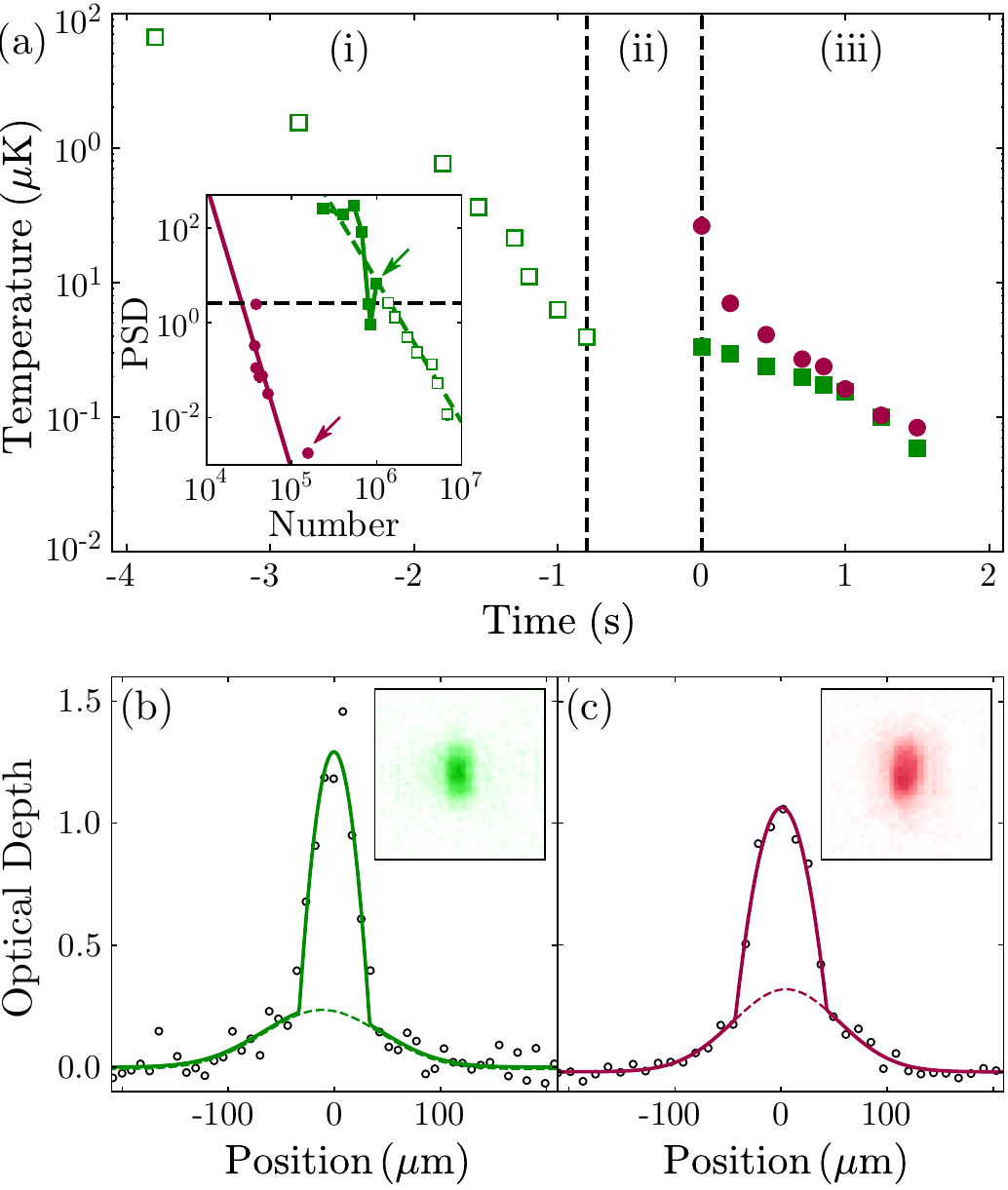}
\caption{Evaporation of Cs and ${}^{174}$Yb to dual-degeneracy in the bichromatic optical dipole trap (BODT).  (a) Temperature versus time during the initial evaporation of the Yb sample (i), preparation of the Cs sample (ii), and subsequent dual-species evaporation sequence (iii). Open (filled) green squares denote $^{174}$Yb before (after) loading Cs into the BODT. Red circles denote Cs. Inset shows phase-space density (PSD) as a function of atom number. In the absence of Cs, the Yb efficiency is $\gamma_{\mathrm{Yb}} = 3.0(6)$ [fit to open green squares shown by the dashed green line]. The solid red line with $\gamma = 6$, and solid green line connecting the Yb data points are guides to the eye for the dual-evaporation of Cs and Yb. Arrows denote initial dual-evaporation measurement at $t = 0$\,s, which occurs pre-thermalization. (b) and (c) Cross-sections through the center of the optical depth (OD) images (inset) for the $^{174}$Yb and Cs BECs, respectively.  Solid lines are a bimodal fit to the 1D OD cuts. Dashed lines show the Gaussian fit to the thermal contribution.  Condensate atom numbers extracted from the fits are $N_\mathrm{Yb} = 6 \times 10^4$ and $N_\mathrm{Cs}  = 1.6 \times 10^4$ with an Yb (Cs) condensate fraction of 0.5 (0.4).
	\label{fig:PSD}}
\end{figure}
\begin{figure}[t]
		\includegraphics[width=0.95\linewidth]{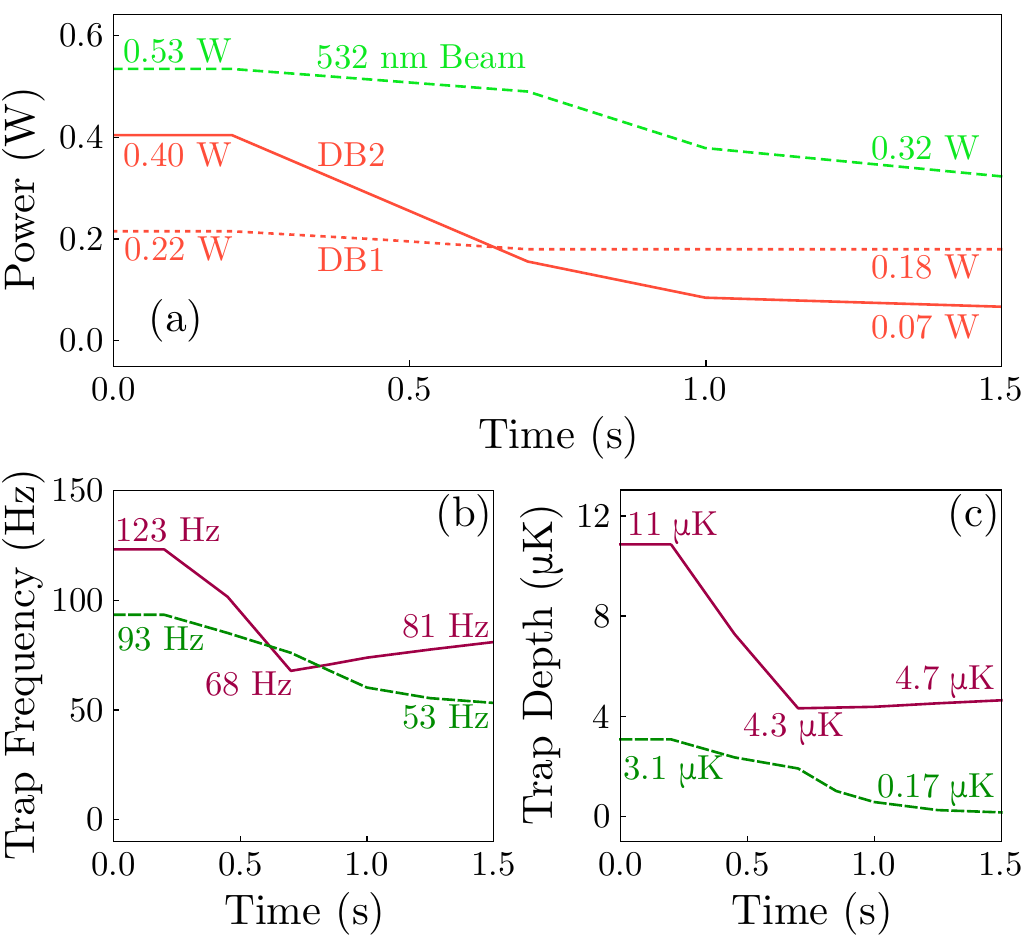}
\caption{Timing sequence for dual-species evaporation in the bichromatic optical dipole trap (BODT). Forced evaporation of Yb is used to sympathetically cool Cs, resulting in dual-degeneracy. (a) Beam powers for dimple beam 1 (DB1, dotted red line), dimple beam 2 (DB2, solid red line), and the 532$\,$nm beam (dashed green line). (b) Geometric mean trap frequencies for Cs (solid red line) and Yb (dashed green line). (c) Trap depths for Cs (solid red line) and Yb (dashed green line). Trap depths include tilting of the trap due to gravity. 
	\label{fig:timing2}}
\end{figure}

At the beginning of the dual-evaporation procedure ($t = 0\,\mathrm{s}$) we have $1.6 \times 10^5$ Cs atoms at a temperature of $2.6 \, \mu$K, and $9.9 \times 10^5$ Yb atoms at a temperature of 240\,nK colocated in the BODT.  Figure~\ref{fig:PSD}(a, iii) shows the evolution of the temperature of the Cs (red circles) and Yb (filled green squares) samples during the dual-evaporation process. For comparison, we also show the Yb only evaporation ramp, from the preparation of the Yb sample, as open green squares in Fig.~\ref{fig:PSD}(a, i).  The corresponding PSD versus atom number data is plotted in the inset of Fig.~\ref{fig:PSD}(a).   Arrows point to the PSD data points corresponding to $t = 0\,\mathrm{s}$, where Yb has just crossed the BEC phase transition (horizontal dashed line at PSD = 2.61), and the mixture has not yet thermalized. The trap powers are held constant for an initial 200\,ms thermalization period, which leads to a factor of 20 increase in the Cs PSD and a concomitant factor of 7 decrease in the Yb PSD. The impact on the Yb PSD is lessened due to the number imbalance in the mixture. The moderately large interspecies scattering length $a_\mathrm{CsYb} = -75 \, a_{0}$ is highly favourable for both thermalisation and sympathetic cooling.
 
At $t = 200\,$ms, we begin to cool the mixture by forced evaporation of the Yb atoms. As shown in Fig.~\ref{fig:timing2}(a), we reduce the power of the $532 \,$nm beam to reduce the Yb trap depth, and modify the powers of the $1070 \,$nm beam during the evaporation for optimal Cs cooling. Throughout the dual-evaporation process we use a bias field of $B_0 = 22.2 \, \mathrm{G}$  $(a_\mathrm{Cs} = 275\,a_0)$ to place the Cs intraspecies scattering length in the vicinity of the Efimov minimum in the three-body recombination rate \cite{Kraemer2006}. We find similar Cs performance for bias fields within the ranges 17 to 19.5\,G and 20.5 to 25\,G so long as we avoid the Cs Feshbach resonance at 19.9 G. Figures~\ref{fig:timing2}(b) and \ref{fig:timing2}(c) show the corresponding mean trap frequency and trap depth, respectively, for both Cs (solid red lines) and Yb (dashed green lines). At the end of the BODT power ramps, the mixture is confined in a trap with frequencies $(\omega_\mathrm{x},\omega_\mathrm{y},\omega_\mathrm{z})/2\pi = (10,120,80)$\,Hz for Yb and $(\omega_\mathrm{\tilde{x}},\omega_\mathrm{\tilde{y}},\omega_\mathrm{z})/2\pi = (40,70,260)$\,Hz for Cs, with typically $\sim$ 10\% uncertainties. The axes $\tilde{x}$ and $\tilde{y}$ for the final Cs trap are rotated $20^{\circ}$ from the Yb trap axes.

The evaporation trajectories are illustrated by the PSD versus atom number plots [Fig.~\ref{fig:PSD}(a), inset]. We find the efficiency $\gamma = d [\ln (\mathrm{PSD})] / d [\ln (N)]$ for the Yb evaporation prior to Cs loading to be $\gamma_{\mathrm{Yb}} = 3.0(6)$ (dashed green line fit to the open squares). Yb recrosses the BEC transition early in the dual-evaporation sequence. Sympathetic cooling of Cs continues after Yb crosses the phase transition and remains efficient; the solid red line with $\gamma = 6$ is a guide to the eye. 
Cs crosses the BEC transition towards the end of the dual-species evaporation sequence. We find we can form dual-BECs with relatively large condensate atom numbers up to $N_\mathrm{Yb} \sim 1 \times 10^5$, $N_\mathrm{Cs} \sim 2 \times 10^4$, but the BECs also have high thermal fractions. Figures~\ref{fig:PSD}{(b) and \ref{fig:PSD}(c)} show optical depth (OD) profiles for a representative dual-BEC measured by absorption imaging after a 25 ms time of flight. Once Cs has crossed the BEC transition, sympathetic cooling becomes less efficient.  As a result, when optimizing the dual-evaporation for pure BECs, we typically obtain pure degenerate Bose-Bose mixtures with $4 \times 10^{3}$ to $5 \times 10^{3}$ Cs atoms and $5 \times 10^{4}$ to $7 \times 10^{4}$  $^{174}$Yb atoms. We note that the final atom numbers and BEC purity are highly sensitive to the BODT beam alignment.  However, BEC overlap is aided by the attractive interspecies interaction. The mixture itself is stable against collapse as the repulsive intraspecies interactions stabilize the attractive interspecies interaction.    
\begin{figure}[t]
		\includegraphics[width=0.95\linewidth]{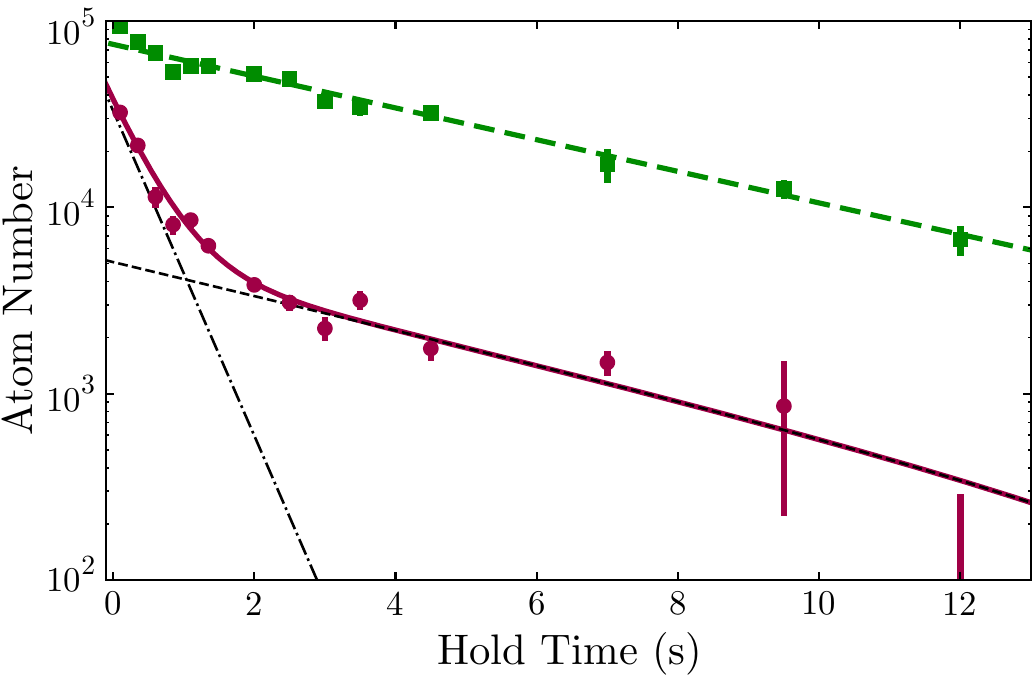}
\caption{Lifetime of dual-degenerate Cs--${}^{174}$Yb mixture in the bichromatic optical dipole trap (BODT).  Plot shows Cs (red circles) and Yb (green squares) atom numbers versus hold time after the final dual-species evaporation ramp. Fit to Cs data points (solid red line) is a double exponential with offset with $\tau_\mathrm{Cs,1} = 0.5(1)$\,s and $\tau_\mathrm{Cs,2} = 5(4)$\,s.  After $1\,\mathrm{s}$ of hold time Cs is no longer a BEC. Fit to Yb data points (dashed green line) is a single exponential with $\tau_\mathrm{Yb} = 5(1)$\,s. Early data points $t < 1.2$\,s are excluded from the fit to the Yb data. 
	\label{fig:life}}
\end{figure}
\begin{figure}[t]
		\includegraphics[width=0.95\linewidth]{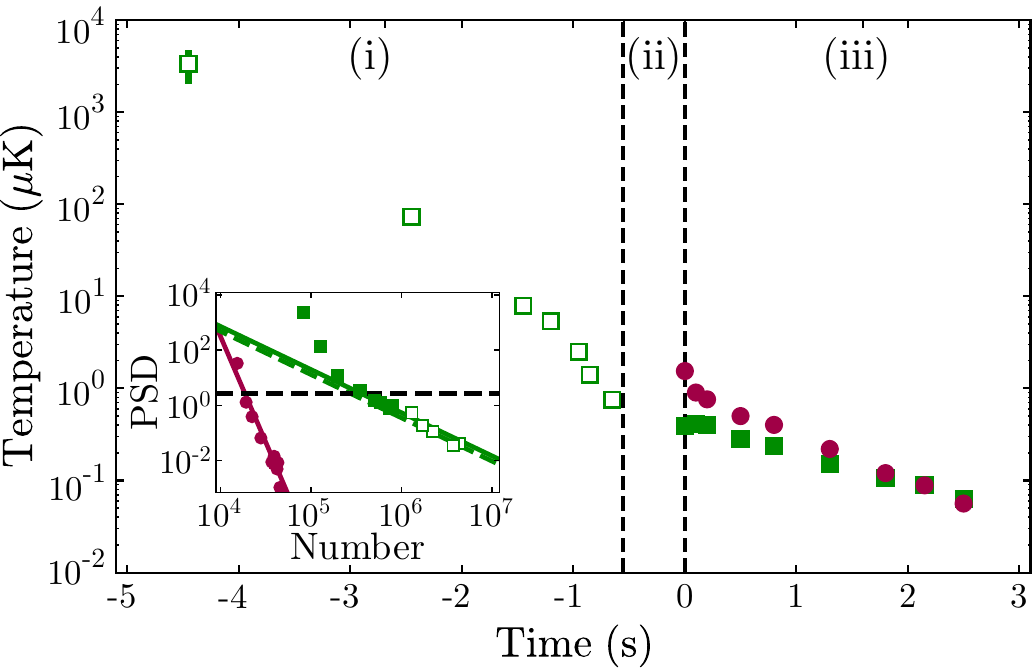}
\caption{Evaporation of Cs and ${}^{170}$Yb to dual-degeneracy in the bichromatic optical dipole trap (BODT). Temperature versus time during the initial evaporation of the Yb sample (i), preparation of the Cs sample (ii), and subsequent dual-species evaporation sequence (iii). Open (filled) green squares denote $^{170}$Yb before (after) loading Cs into the BODT. Red circles denote Cs.  Inset shows phase-space density PSD as a function of atom number. Fits to PSD data give a Cs evaporation efficiency of $\gamma_{\mathrm{Cs}} = 7.7(6)$, Yb efficiency of $\gamma_{\mathrm{Yb}} = 1.6(2)$ with Cs present, and an Yb efficiency of $\gamma_{\mathrm{Yb}} = 1.6(5)$ with Cs absent.  
	\label{fig:PSD170Yb}}
\end{figure}
\begin{figure}[ht!]
		\includegraphics[width=0.95\linewidth]{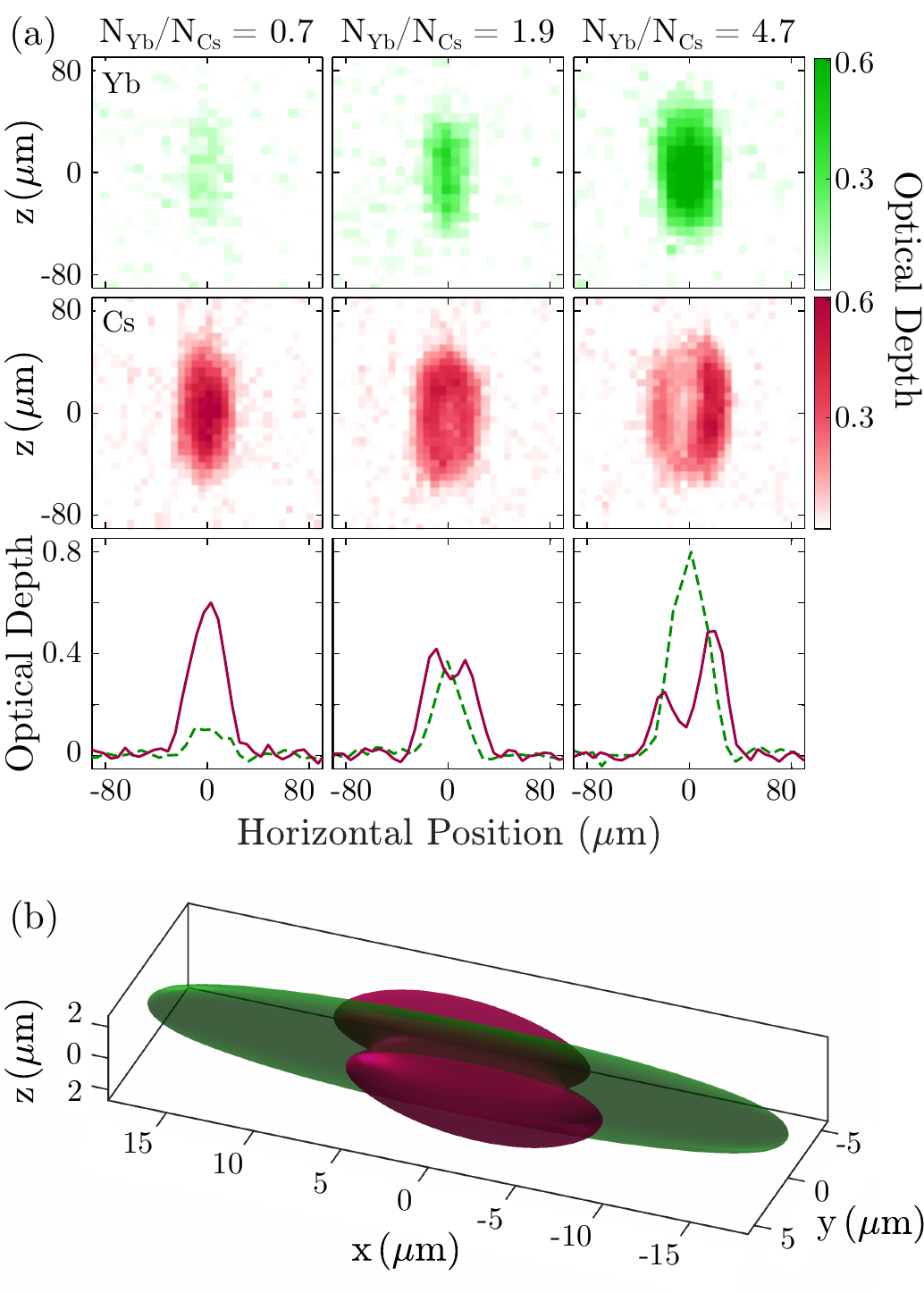}
\caption{Number dependent immiscibility in a dual-degenerate mixture of Cs and $^{170}$Yb. (a) Optical depth (OD) profiles showing $^{170}$Yb (top row) and Cs (middle row). OD profiles are extracted from absorption images taken after 20\,ms time of flight expansion. The bottom row shows horizontal cross sections through the center of the OD profiles (averaged over a range of 6 pixels centered on $\Delta z = 0$) for Cs (solid red lines) and Yb (dashed green lines).  We have independently centered each image, using a Gaussian fit to the atom cloud.  From left to right the atom number ratios are $N_\mathrm{Yb} /  N_\mathrm{Cs} = 0.7$,  $N_\mathrm{Yb} /  N_\mathrm{Cs} = 1.9$, and $N_\mathrm{Yb} /  N_\mathrm{Cs} = 4.7$, with $N_\mathrm{Cs} = 5.5 \times 10^3$  for all three images. Scattering lengths are $a_\mathrm{Yb} = 64 \, a_0$, $a_\mathrm{Cs} = 440 \, a_0$, and $a_\mathrm{CsYb} = 96 \, a_0$.  Note that the condensates shown in the two rightmost columns are partially immiscible which is observable in the distortions of the Cs OD profile.  The slight asymmetry observed in the rightmost Cs profile is due to a small misalignment in horizontal trap centers.  (b) Ground state profiles for Cs (red: double-lobed) and Yb (green: cigar-shaped) calculated using imaginary time propagation of the coupled Gross-Pitaevski equations. The choice of atom number $N_\mathrm{Yb} = 2.6 \times 10^4$, $N_\mathrm{Cs} = 5.5 \times 10^3$ corresponds to the rightmost experimental images, and further confirms the partial immiscibility observed for Cs-$^{170}$Yb mixtures with larger Yb atom numbers.\label{fig:Cs170Yb}}
\end{figure}

It is possible to vary the atom number ratio of the mixture by tuning the Cs MOT loading \footnote{we load to a threshold MOT fluorescence value to maintain consistency in the atom number from shot-to-shot}. However, we find we are limited to mixtures with a number imbalance in favour of Yb. The large Cs three-body recombination rate limits the number of Cs atoms that can be produced at low temperatures for a given trapping potential. Loading larger Cs atomic samples significantly increases the heat load on Yb, resulting in poor Yb evaporation efficiency and does not significantly increase the Cs condensate numbers.  We note that if more balanced atom numbers are needed, Yb atoms can be selectively removed after the full evaporation sequence using a pulse of the $399 \, \mathrm{nm}$ light used for imaging on the $^1{S}_0$ to $^1{P}_1$ transition. 

We measure the lifetime of the dual-degenerate mixture by holding the gas in the final BODT. Figure.~\ref{fig:life} shows the number of atoms for both species versus time.  The plot shows a double-exponential decay for the Cs atom number loss (solid red line) and a single exponential decay for Yb (dashed green line). We extract Cs 1/e lifetimes of $\tau_\mathrm{Cs,1} = 0.5(1)\,\mathrm{s}$ and $\tau_\mathrm{Cs,2} = 5(4)\,\mathrm{s}$. We find similar Yb behaviour at long hold times with an Yb 1/e lifetime of $\tau_\mathrm{Yb} =  5(1)\,\mathrm{s}$. Early data points $t < 1.2\,$s are excluded from the single-exponential fit to the Yb data. We note that the Cs atoms are no longer condensed after the first $1\,$s of hold time, while the Yb atoms remain condensed throughout.

The actual loss mechanisms are difficult to untangle. The atoms experience heating and loss due to three-body recombination, as well as radiative heating due to photon scattering and technical heating in the BODT.  We observe a heating rate for the thermal Cs atoms of $19(1)$\,nK/s over the time period from 2 to 8 s, which is of the same order of magnitude as the predicted radiative heating rate for Cs at the final BODT beam powers of 26\,nK/s.  We expect any Cs heating to be partially mitigated by thermal contact with the Yb BEC.  

We use the short-time Cs 1/e lifetime $\tau_\mathrm{Cs,1} = 0.5 \, \mathrm{s}$ together with numerically calculated condensate densities $n_\mathrm{Cs}$ and $n_\mathrm{Yb}$ to place an upper bound on the potential Cs three-body loss rate $K_\mathrm{3,Cs} = (\tau_\mathrm{Cs,1} \langle n_\mathrm{Cs} n_\mathrm{Cs}\rangle_\mathrm{sp})^{-1}$ and Cs-Yb three-body loss rate $K_\mathrm{3,CsYb} = (\tau_\mathrm{Cs,1} \langle n_\mathrm{Cs} n_\mathrm{Yb}\rangle_\mathrm{sp})^{-1}$ \cite{Cho2011}. Here $\langle \rangle_\mathrm{sp}$ denotes a 3D spatial average bounded by the extent of the Cs cloud. We compute the ground state 3D atom densities numerically using imaginary time propagation of the full 3D coupled Gross-Pitaevskii equations with a conservative estimate for the condensate numbers of $N_\mathrm{Cs} = 5 \times 10^{3}$, $N_\mathrm{Yb} = 5 \times 10^{4}$, and with $a_\mathrm{Cs} = 275\,a_0$.  We note that the numerically calculated peak atom densities are similar $n_\mathrm{0,Yb} = 2.5 \, n_\mathrm{0,Cs}$ despite the factor of 10 number imbalance in favor of Yb.  This is due to our trapping geometry where Cs experiences a tighter trap.
If we assume that all the atom loss during the first  1\,s of hold time is due to Cs three-body recombination, we find a condensate Cs three-body loss rate of $\sim 1.3 \times 10^{-27} \mathrm{cm}^6/\mathrm{s}$. The thermal three-body loss rate is a factor of 3! larger giving $K_\mathrm{3,Cs} \sim 8 \times 10^{-27} \mathrm{cm}^6/\mathrm{s}$, in line with previous measurements \cite{Guttridge2017, Kraemer2006}. Alternatively, if we assume that all the loss is accounted for by Cs--Yb three-body recombination, we place an upper bound on the Cs--Yb three body loss process of $K_\mathrm{3,CsYb} \sim 2.4 \times 10^{-27} \mathrm{cm}^6/\mathrm{s}$.  In practice both intraspecies and interspecies three body loss processes may play some role, and it is difficult to decouple the two.

\section{Dual-Species evaporation with Cs and \textsuperscript{170}Yb}\label{sec:170Yb}
We report a dual-degenerate Bose-Bose mixture of Cs and $^{170}$Yb using a similar evaporation sequence to that of Cs and $^{174}$Yb discussed in Sec.~\ref{sec:BODT2} and Sec.~\ref{sec:evap}.  The Cs--$^{170}$Yb interspecies scattering length $a_\mathrm{CsYb} = 96 \, a_0$ and the $^{170}$Yb intraspecies scattering length $a_\mathrm{Yb} = 64 \, a_0$ are of the same order of magnitude as the Cs and $^{174}$Yb  scattering lengths with one critical difference.  The Cs--$^{170}$Yb interspecies scattering length is positive, leading to a repulsive interspecies interaction, in contrast to the attractive Cs--$^{174}$Yb interaction. Figure~\ref{fig:PSD170Yb} shows the temperature evolution of both Cs and $^{170}$Yb throughout the evaporation process. While the dual-evaporation scheme largely follows that of $^{174}$Yb, we note that initial preparation of the $^{170}$Yb sample takes longer. This is due to a combination of the lower natural abundance (3\% for $^{170}$Yb compared to 32\% for $^{174}$Yb) requiring longer MOT loading times, and a longer single species evaporation stage due to the smaller intraspecies scattering length. The Cs evaporation efficiency is $\gamma_\mathrm{Cs} = 7.7(7)$, as shown by the fits to PSD versus atom number in the inset of  Fig.~\ref{fig:PSD170Yb}, with Yb evaporation efficiencies of $\gamma_\mathrm{Yb} = 1.6(5)$ for Yb alone and $\gamma_\mathrm{Yb} = 1.6(2)$ for Yb with Cs. Typical atom numbers for pure Cs--$^{170}$Yb dual-BECs are $N_\mathrm{Yb} \sim 4 \times 10^4$, and $N_\mathrm{Cs} \sim 1 \times 10^4$.  

In degenerate mixtures with repulsive interactions, the relevant instability is a phase-instability. This manifests itself as the transition between the two BECs being miscible or immiscible depending on the balance between interspecies and intraspecies interaction energies \cite{Papp2008, Wang2015}.  In the limit of homogeneous density distributions within the Thomas-Fermi regime, the miscible/immiscible transition is characterised by $\Delta = g_{11}g_{22}/g_{12}^2 - 1$  with phase separation occurring when $\Delta \leq 0$ \cite{Pethick2008}. Here
\begin{equation}
g_{ij} = \frac{2\pi\hbar^2 a_{ij} (m_i + m_j)}{m_i m_j},
\end{equation}
is the interaction coupling constant. For atoms confined within a harmonic trap, the density profile is inhomogenous, however $\Delta$ still provides a reasonable measure of the miscibility so long as the mixture is balanced ($N_1 \sim N_2$). Under this assumption our mixture would become immiscible for $a_\mathrm{Cs} \leq 147\,a_0$. However, for imbalanced mixtures such as ours trapping parameters and relative atom numbers play a significant role \cite{Lee2016b}.  
Figure~\ref{fig:Cs170Yb}(a) shows OD profiles for pairs of Yb (top row) and Cs (middle row) BECs with varying atom number ratios: from left to right $N_\mathrm{Yb} /  N_\mathrm{Cs} = 0.7$,  $N_\mathrm{Yb} /  N_\mathrm{Cs} = 1.9$, and $N_\mathrm{Yb} /  N_\mathrm{Cs} = 4.7$. The Cs atom number is fixed at $N_\mathrm{Cs} = 5.5 \times 10^3$ for all three panels.  The bottom row of Fig.~\ref{fig:Cs170Yb}(a) shows horizontal cross sections through the center of the OD profiles (averaged over a range of 6 pixels centered on $\Delta z = 0$) for Cs (solid red lines) and Yb (dashed green lines). We have centered each image independently, using a Gaussian fit to the atom cloud.  Here we have formed a BEC at the usual 22.2 G bias field, and then adiabatically ramped the bias field to 26.0\,G ($a_\mathrm{Cs} = 440\,a_0$), yielding $\Delta = 2$ comfortably above the homogeneous miscibility transition.  Remarkably, we observe a shift towards the immiscible regime as the number of Yb atoms increases.  Note the reduction of Cs atoms in the center of the BEC image shown in the bottom-center and bottom-right panels of Fig.~\ref{fig:Cs170Yb}(a). This shift towards immiscibility as the ratio $N_\mathrm{Yb}/N_\mathrm{Cs}$ increases is likely due to competing roles of the mis-matched trapping potentials and atom number imbalance \cite{Lee2016b}. 
We note that BEC overlap with the repulsive Cs--$^{170}$Yb interspecies interaction is much more sensitive than for the attractive case where the interaction acts to pull the BECs towards each other.  We observe very different profiles if the clouds are slightly offset as indicated by the asymmetry observed in the rightmost Cs profile in Fig.~\ref{fig:Cs170Yb}(a).  In the extreme case, we observe all of the Cs atoms pushed over into a single lobe. This motivates moving towards a `special-wavelength' ODT (see discussion in Sec.~\ref{sec:BODT}) following BEC production to both bypass alignment issues inherent in the BODT, and better match the Cs and Yb trap frequencies.

For comparison, we compute the ground state BEC density profiles numerically using imaginary time propagation of the full 3D  coupled Gross-Pitaevskii equations \cite{Pethick2008}.  Figure~\ref{fig:Cs170Yb}(b) shows representative ground state BEC density profiles for $N_\mathrm{Yb} = 2.6 \times 10^4$, $N_\mathrm{Cs} = 5.5 \times 10^3$, and $a_\mathrm{Cs} = 440 \, a_0$ corresponding to the rightmost panels of Fig.~\ref{fig:Cs170Yb}(a). The numerical simulations confirm that the BECs are only partially miscible with reduced Cs density in the central region of overlap. The lines of sight and low resolution of the Cs and Yb imaging systems in our current experimental apparatus wash out most of this effect, but this is something to be explored further in future work after upgrades to the imaging systems.

\section{Conclusion}\label{sec:conclusion}
In summary, we have employed a bichromatic optical dipole trap operating at 532\,nm and 1070\,nm to produce the first quantum mixtures of pure Cs--$^{174}$Yb BECs with typical atom numbers $N_\mathrm{Yb} \sim 5 \times 10^4$ and $N_\mathrm{Cs} \sim 5 \times 10^3$, and pure Cs--$^{170}$Yb BECs with typical atom numbers $N_\mathrm{Yb} \sim 4 \times 10^4$ and $N_\mathrm{Cs} \sim 1 \times 10^4$. Our dual-evaporation scheme takes advantage of the favorable interspecies interactions to enable efficient sympathetic cooling of Cs by either $^{174}$Yb with the attractive interspecies scattering length $a_\mathrm{CsYb} = -75 \, a_0$, or  $^{170}$Yb with the repulsive interspecies scattering length $a_\mathrm{CsYb} = 96 \, a_0$. These Cs--Yb mixtures provide a versatile system for studies of impurity physics, collective dynamics, binary fluid dynamics, and two-component quantum turbulence. Of particular relevance to future studies is the fact that Cs and Yb may be perturbed independently of each other with Cs-blind optical potentials and Yb-blind magnetic potentials. Furthermore, tuning the Cs scattering length allows fine control over the balance of the mean-field contributions. Further upgrades to our apparatus such as a high-resolution imaging system and a tunable optical potential ($\lambda \sim 460\,\mathrm{nm}$) will enable future work focused on the collective behaviour of these mixtures including exploring the quantum droplet / BEC phase transition \cite{Petrov2015, Cabrera2018, Semeghini2018} and the miscible / immiscible phase transition \cite{Papp2008, Wang2015}.

The data presented in this paper are available from \cite{WilsonData}.

\begin{acknowledgments}
We thank Jeremy Hutson, Matthew Frye, Tom Billam, Nick Parker, Nick Proukakis, and I-Kang Liu for helpful discussions. We acknowledge support from the UK Engineering and Physical Sciences Research Council (grant numbers EP/P01058X/1 and EP/T015241/1). 
\end{acknowledgments}

\appendix*
\section{Quadrant Photodiode Setup}
\begin{figure}[b]
		\includegraphics[width=0.95\linewidth]{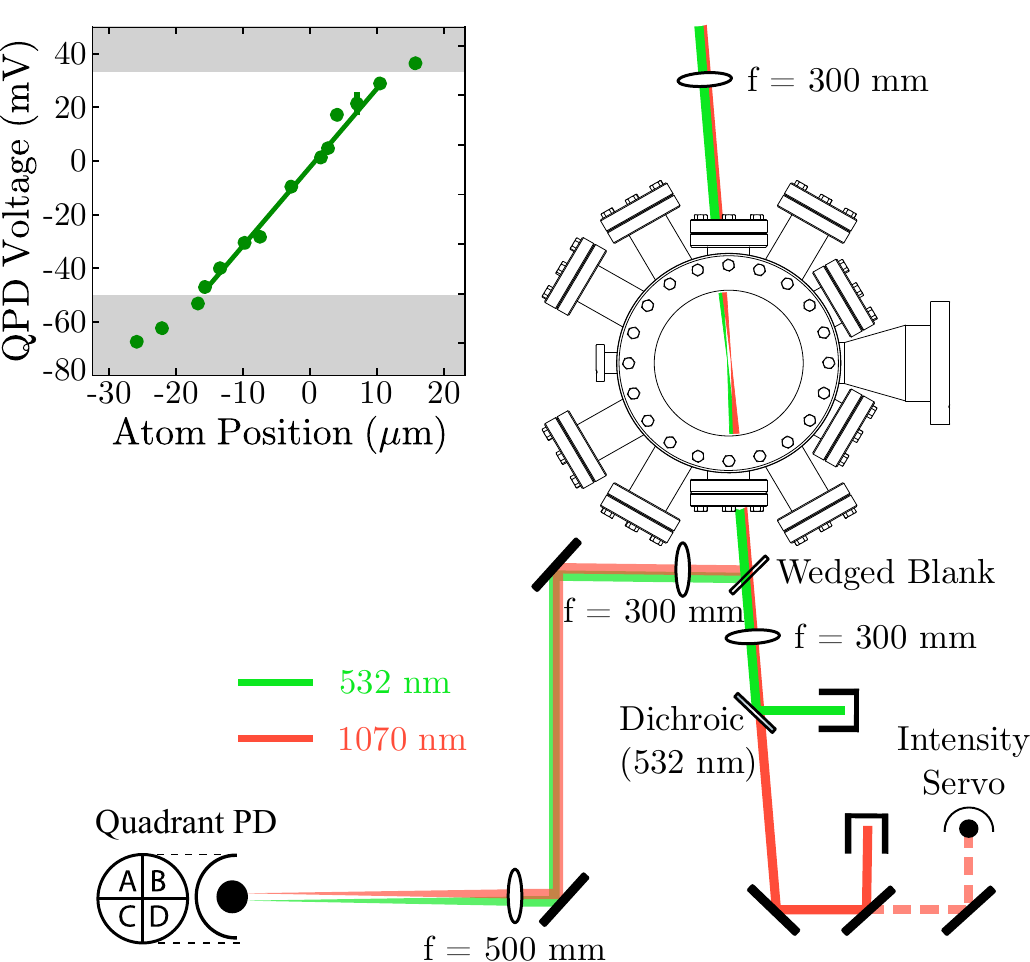}
\caption{Optical setup to track the relative beam positions of the copropagating 532\,nm and 1070\,nm beams that make up the bichromatic optical dipole trap (BODT).  The beam positions in the plane of the atoms are imaged onto a quadrant photodiode (QPD) using a pair of achromatic lenses with focal lengths $f = 300$\,mm and $f = 500$\,mm. Inset: calibration of the QPD sensitivity for the 532\,nm beam.  The measured vertical QPD voltage versus the vertical center of mass (CoM) Yb atom position is plotted in green circles. Here the vertical CoM position of a cloud of Yb atoms confined solely by the 532\,nm beam gives an independent measure of the 532\,nm beam position.   Grey shaded areas denote QPD voltages excluded from the linear fit (green line), which gives a sensitivity of $2.9(1) \, \mathrm{mV}/\mu \mathrm{m}$.  
	\label{fig:QPD1}}
\end{figure}

Figure~\ref{fig:QPD1} shows the optical setup used to monitor the relative
positions of the copropagating 532\,nm and 1070\,nm (DB1) beams. The dipole trap beams in the plane of the BEC are imaged onto a quadrant photodiode (QPD) using a pair of achromatic lenses, with focal lengths of 300\,mm and 500\,mm, chosen to minimise the shift in the object plane between the two wavelengths.  To track the relative position of the two beams, at the end of an experimental run we image the beams sequentially, illuminating the QPD first with DB1 and then with the 532\,nm beam.

We use a First Sensor QPD mounted in an evaluation board (QP50-6-18u SD2), which has a narrow 18\,$\mu$m gap between the photodiodes. The evaluation board provides voltage outputs for the horizontal $V_\mathrm{H}$ and vertical $V_\mathrm{V}$ beam positions with $V_\mathrm{H}$ the difference between left and right halves of the photodetector, and $V_\mathrm{V}$ the difference between top and bottom halves of the photodetector. For a Gaussian beam, and a narrow gap QPD, the QPD voltage as a function of beam position has the form of an error function, which can be approximated as linear in the central region of the QPD [see inset of Fig.~\ref{fig:QPD1}]. 
We calibrated the sensitivity of the QPD for the 532\,nm (1070\,nm) beam by plotting the vertical QPD voltage versus the vertical center of mass (CoM) position of Yb atoms confined solely by the $532 \, \mathrm{nm}$ ($1070 \, \mathrm{nm}$) beam. Here the vertical CoM position of a cloud of Yb atoms confined in a single beam gives an independent measure of that beam's vertical position at the location of the atoms.  We measure vertical sensitivities of 2.9(1)\,mV/$\mu$m and 2.3(1)\,mV/$\mu$m for the 532\,nm and 1070\,nm beams respectively. The inset of Fig.~\ref{fig:QPD1} shows the calibration of the QPD sensitivity for the 532\,nm beam.  The\,mV scale is set by the optical power incident on the QPD and the response of the QPD at the two wavelengths; $V_\mathrm{Sum} = V_\mathrm{H} + V_\mathrm{V} \simeq 200$\,mV for the 1070\,nm beam and $V_\mathrm{Sum} \simeq 100$\,mV for the 532\,nm beam. We are unable to perform the same calibration for the horizontal direction given the relative alignment of the copropagating BODT beams and the Cs and Yb imaging systems (see Fig.~\ref{fig:beams}), therefore we use the vertical calibration as a reasonable estimate for the horizontal calibration as well. 

%

\end{document}